\documentclass{mn2e} 
\usepackage{times, xspace}
\input{psfig.sty}

\newcommand\rhoelse{$\rho_{\rm disk+irr}$\xspace}
\newcommand\rhosph{$\rho_{\rm sph}$\xspace}
\newcommand\rhostar{$\rho_*$\xspace}
\newcommand\rhobh{$\rho_{\rm BH}$\xspace}

\newif\ifAMStwofonts

\title[Tracing the cosmological assembly of stars and SMBH]
      {Tracing the cosmological assembly of stars and supermassive
      black holes in galaxies}

\author[Merloni, Rudnick \& Di Matteo]   {Andrea Merloni, Gregory
  Rudnick \&  Tiziana Di Matteo 
\\Max-Planck-Institut f\"ur Astrophysik,
Karl-Schwarzschild-Strasse 1, D-85741, Garching, Germany }

\date{}

\begin{document}

\maketitle

\label{firstpage}

\begin{abstract}
  We examine possible phenomenological constraints for the joint
  evolution of supermassive black holes (SMBH) and their host spheroids.  We
  compare all the available observational data on the redshift
  evolution of the total stellar mass and star formation rate density
  in the Universe with the mass and accretion rate density evolution
  of supermassive black holes, estimated from the hard X-ray
  selected luminosity function of quasars and active galactic nuclei
  (AGN) for a given radiative efficiency, $\epsilon$.  We assume that
  the ratio of the stellar mass in spheroids to the black hole mass
  density evolves as $(1+z)^{-\alpha}$, while the ratio of the stellar
  mass in disks~$+$~irregulars to that in spheroids evolves as
  $(1+z)^{-\beta}$, and we derive constraints on $\alpha$, $\beta$ and
  $\epsilon$.  We find that $\alpha>0$ at the more than 4-sigma level,
  implying a larger black hole mass at higher redshift for a given
  spheroid stellar mass.  The favored values for $\beta$ are typically
  negative, suggesting that the fraction of stellar mass in spheroids
  decreases with increasing redshift. This is consistent with recent
  determinations that show that the mass density at high redshift is
  dominated by galaxies with irregular morphology. In agreement with
  earlier work, we constrain $\epsilon$ to be between 0.04 and 0.11,
  depending on the exact value of the local SMBH mass density, but
  almost independently of $\alpha$ and $\beta$.
\end{abstract}


\section{Introduction}

Recent work has shown that supermassive black holes (SMBH) are
ubiquitous in the centers of nearby galaxies.  The observational
evidence indicates that the mass of the central black hole is
correlated with spheroid luminosity and mass (e.g., Kormendy and
Richstone 1995; Magorrian et al. 1998) and also with the velocity
dispersion of the spheroid (Ferrarese \& Merritt 2000; Gebhardt et al.
2000; Tremaine et al. 2002), suggesting that the process that leads to
the formation of galactic spheroids must be intimately linked to the
growth of the central SMBH.  In addition, studies of active galactic
nuclei (AGN) hosts in the local universe with the Sloan Digital Sky
Survey (SDSS) have also shown that AGN activity is closely related to
star formation in local galaxies (Kauffmann et al. 2003; Heckman et
al. 2004).  From a theoretical perspective, several groups have
attempted to investigate the link between the cosmological evolution
of QSOs and formation history of galaxies within the context of
semi-analytic and numerical models of galaxy formation and evolution
(e.g., ~Monaco et al.~2000; Kauffmann \& Haehnelt~2000; Wyithe \&
Loeb~2003; Di Matteo et al. 2003; Granato et al.~2004; Haiman, Ciotti
\& Ostriker 2004).
 
In this paper we make a phenomenological investigation of the link
between the growth of black holes and the buildup of their spheroidal
hosts over time. Specifically, we compare the evolution of the black
hole mass density, \rhobh, to that of the observed mass density of
stars in galaxies, \rhostar, and derive constraints on their
co-evolution.

To derive \rhobh as a function of redshift, we use the recently
determined mass density of local black holes (Yu \& Tremaine 2002;
Aller \& Richstone 2002; Marconi et al. 2004; Shankar et al. 2004),
and evolve it back to $z \sim 3$ using the observed AGN X-ray
luminosity functions (see also Merloni 2004 and references therein). 
The black hole assembly
history gathered in such a way assumes that the majority of black hole
growth is due to mass accretion and hence is a function of the efficiency
associated with mass to (radiative) energy conversion. This provides
directly the evolution of mass accretion rate density as well, given
by $\dot{\rho}_{\rm BH}\equiv \Psi_{\rm BH}$.

To constrain the build-up of stars in the Universe we use the
independent measurements of the total star formation rate density
($\Psi_{*}$) and the total stellar mass density (\rhostar)
evolution.  Because current observations indicate that the black hole
mass in galaxies is related to the spheroidal component, we will split
up the observed \rhostar into two terms.  We will hereafter refer to
the mass density of stars in spheroids and disks~$+$~irregulars as
\rhosph and \rhoelse respectively. In addition to counting the mass
contribution from pure spheroid, disk, and irregular galaxies, these
terms also include masses in the spheroid and disk components of
individual galaxies.  Estimates of the ratio of \rhoelse to \rhosph at
$z=0$ differ by more than a factor of 5, ranging from 0.2-0.3 (Salucci
\& Persic 1999; Fukugita, Hogan \& Peebles 1998; Fukugita \& Peebles
2004) to 1.0-1.2 (Schechter \& Dressler 1987; Benson et al. 2002). The
redshift evolution of this ratio should depend not only on the rate of
star formation in galaxies of different morphologies at different
epochs, but also on the role of mergers, interactions, and secular
instabilities in determining galaxy morphology. Both from the
theoretical and observational point of view, such an evolution has
been so far very hard to quantify (see e.g. Baugh, Cole \& Frenk 1996;
van den Bosch 1998; Brinchmann \& Ellis 2000; Conselice et al. 2004;
Combes 2004).

 The rest of the paper is organized as follows. In Section 2 we
 briefly summarize and describe recent measurements of \rhostar and
 $\Psi_*$ as functions of redshift. In Section 3 we derive the black
 hole mass density evolution using the X-ray luminosity function and
 the local black hole mass function. Under the assumption that
 $\rho_{\rm sph}(z)\propto \rho_{\rm BH}(z)(1+z)^{-\alpha}$ and that,
 at the same time, ${\rho_{\rm disk+irr}}/{\rho_{\rm sph}}$ evolves as
 $(1+z)^{-\beta}$, we then compare the history of black hole and total
 stellar mass assembly in the Universe and derive our global
 constraints on $\epsilon$, $\alpha$, and $\beta$ (Section 4). Finally,
 in Section 5 we summarize and discuss our results and explore their
 implications.

 Throughout this paper, we adopt a background cosmological model in
 accordance with the Wilkinson Microwave Anisotropy Probe ({\it WMAP})
 experiment. The model has zero spatial curvature, a cosmological
 constant, $\Omega_{\Lambda} =0.71$ a Hubble constant $H_0 = 72$
 km s$^{-1}$, dominated by cold dark matter with $\Omega_{m} = 0.29$
 and $\Omega_b =0.047$ (Spergel et al. 2003).

\section{The stellar mass density evolution}

 In this work we make use of all available measurements of the stellar
 mass and star formation rate densities from the literature.  The
 stellar mass density, \rhostar, has been measured in the redshift
 range from $z=0$ to $z\sim 3$ from surveys that select galaxies by
 rest-frame optical or Near Infrared (NIR) light and, in general,
 estimate the stellar masses and mass-to-light ratios of individual
 galaxies by fitting spectral synthesis models to optical-NIR spectral
 energy distributions (see e.g. Brinchmann \& Ellis 2000; Cole et
 al. 2001; Dickinson et al. 2003; Fontana et al. 2003; Drory et
 al. 2004). One exception is Rudnick et al. (2003) who have determined
 the evolution in \rhostar from the change in the luminosity density
 and the global mass-to-light ratio, the latter being determined by
 using simple models to interpret the cosmically averaged rest-frame
 optical colors. Although cosmic variance is a dominant source of
 error for all but the local values, all support a scenario in which
 \rhostar at $z\sim 3$ was a factor of $\sim 7-20$ times lower that it
 is today and that roughly $50\%$ of the current stellar mass was
 assembled by $z=1-1.5$.

 The measurements of \rhostar as a function of redshift also match
 well with the independently measured integrated global star formation
 rate density (SFR). The SFR measurements are usually obtained from
 e.g., optical emission line measurements, MIR dust luminosities,
 sub-mm observations, or observations of the rest-frame ultraviolet
 light. These typically show a rapid rise of the SFR($z$) out to $z\la
 1.5$ where it peaks, followed by a roughly constant level out to at
 least $z\sim 4$ (see caption of Figure~\ref{fig:fits} for detailed
 references).

 For the most part, not enough information exists to empirically
 divide \rhostar and the SFR into different morphological
 components. However, when comparing these data with \rhobh (see
 below), it is necessary to take into account the fact that local
 black hole masses only correlate with the properties of their
 spheroidal hosts. We account for this by defining a parameter
 $\lambda(z)$ as the ratio of the mass in disks~$+$~irregulars to that
 in spheroids, in the following way:

\begin{equation}
\rho_{*}(z)=\rho_{\rm sph}(z)+\rho_{\rm disk+irr}(z)=\rho_{\rm
  sph}(z)[1+\lambda(z)].
\label{eq:lambda}
\end{equation}
 We then assume that $\lambda(z)$ evolves simply according to
  $\lambda(z)=\lambda_0 (1+z)^{-\beta}$, where $\lambda_0$
 is the value of the disk to spheroid ratio
 in the local universe\footnote{Since the stellar
 mass budget at low redshift is overwhelmingly dominated by spheroids
 and disks, $\rho_{\rm disk+irr}(z=0)\approx\rho_{\rm disk}(z=0)$.  At high
 redshift, where irregulars contribute more to the stellar mass
 density, e.g. Brinchmann \& Ellis (2000), $\rho_{\rm
 disk+irr}\approx\rho_{\rm disk}+\rho_{\rm irregular}$.}.

\section{The black hole mass density evolution}

 Under the standard assumption that black holes grow mainly by
 accretion, the cosmic evolution of the SMBH accretion rate and its
 associated mass density can be calculated from the luminosity
 function of AGN: $\phi(L_{\rm bol},z)=dN/dL_{\rm bol}$, where $L_{\rm
 bol}=\epsilon \dot M c^2$ is the bolometric luminosity produced by a
 SMBH accreting at a rate of $\dot M$ with a radiative efficiency
 $\epsilon$. In practice, the accreting black hole population is
 always selected through observations in specific wavebands.  Crucial
 is therefore the knowledge of two factors: the completeness of any
 specific AGN survey, and the bolometric correction needed in order to
 estimate $L_{\rm bol}$ from the observed luminosity in any specific
 band.

 In recent years, enormous progress has been made in understanding
 both of the above issues. Here we refer to the discussion of Marconi
 et al. (2004), who carry out a detailed comparison between optically
 selected QSOs from the 2dF survey (Boyle et al. 2000), soft X-ray
 (0.5-2 keV) selected AGN (Miyaji, Hasinger and Schmidt 2000), and hard
 X-ray (2-10 keV) selected ones (Ueda et al. 2003), and work out the
 respective bolometric corrections.
 
 The general picture that emerges is that the 2-10 keV AGN luminosity
 function probes by far the largest fraction of the whole AGN
 population. This is primarily due to the fact that hard X-rays are
 less affected by obscuration than either soft X-rays or optical light
 (Marconi et al. 2004).  It has also been shown that most of the
 obscured sources, which are the major contributor to the X-ray
 background light, (see e.g. Hasinger 2003, and references therein;
 Fiore 2003) have typically lower luminosity and lie at a lower
 redshift than the the optically selected QSOs. Therefore, the growth
 of SMBH (and of their associated spheroidal hosts) at $z<1$ is best
 estimated by studying the evolution of the hard X-ray emitting AGN.
 At higher redshifts, optical and hard X-ray selection give consistent
 results (see e.g. Hasinger 2003).

 In the following we will indeed assume that the absorption
 corrected hard X-ray luminosity
 function of AGN best describes the evolution of the {\it entire}
 accreting black holes population between $z=0$ and $z \sim 3$. For
 the sake of simplicity, we will adopt for our calculations the
 luminosity dependent density evolution (LDDE) parameterization of the
 (intrinsic) 2-10 keV luminosity function described by Ueda et al. (2003).

The redshift evolution of the SMBH
accretion rate density can then be easily calculated as follows:
\begin{equation}
\label{eq:bhar_z}
\Psi_{\rm BH}(z)=\int_0^{\infty}\frac{(1-\epsilon)L_{\rm bol}(L_{\rm X})}{\epsilon c^2}
\phi(L_{\rm X},z)dL_{\rm X},
\end{equation}
where $L_{\rm X}$ is the X-ray luminosity in the rest-frame 2-10 keV
band, and the bolometric correction function $L_{\rm bol}(L_{\rm X})$
is given by eq. (21) of Marconi et al. (2004), that also takes 
  into account the observed dependence of the optical-to-X-ray ratio
  $\alpha_{ox}$ on luminosity (Vignali, Brandt \& Schneider 2003). This in turn yields for
the SMBH mass density:
\begin{equation}
\label{eq:rhobh_z}
\frac{\rho_{\rm BH}(z)}{\rho_{\rm BH,0}}=1-\int_0^{z}\frac{\Psi_{\rm BH}(z')}{\rho_{\rm BH,0}}\frac{dt}{dz'}dz'.
\end{equation}

For any given $\phi(L_{\rm X},z)$ and bolometric correction, the exact
shapes of $\rho_{\rm BH}(z)$ and $\Psi_{\rm BH}$ then depend only on
two numbers: the local black holes mass density $\rho_{\rm BH,0}$ and
the (average) radiative efficiency $\epsilon$.

\section{The joint evolution of the stellar and black hole mass densities}
\label{sec:results}
Given the derived $\rho_{\rm BH}(z)$ above, we now want to investigate
how the buildup of spheroidal mass in galaxies is related to the build
up of black hole mass over time.  We make the simple hypothesis that
the redshift evolution of the spheroid and SMBH mass densities are
related by the expression:
\begin{equation}
\label{eq:sph}
\rho_{\rm sph}(z)={\cal A}_0 \rho_{\rm BH}(z)
(1+z)^{-\alpha},
\end{equation}
where ${\cal A}_0$ is defined by:
\begin{equation}
\label{eq:A0}
{\cal A}_0= \rho_{\rm sph,0}/ \rho_{\rm BH,0} = \rho_{*,0} / [\rho_{\rm BH,0} (1 + \lambda_0)].
\end{equation}
For all reasonable values of $\rho_{*,0}$, $\rho_{\rm BH,0}$, and
$\lambda_0$, ${\cal A}_0$ is consistent with the expectation from the
$M_{\rm BH}-M_{\rm sph}$ relation in the local universe (Merritt \&
Ferrarese 2001; Mc Lure \& Dunlop 2002; Marconi \& Hunt 2003; H\"aring
\& Rix 2004). Then, from eqs.~(\ref{eq:lambda}) and (\ref{eq:sph}) we
obtain the desired relation between total stellar and black holes mass
densities:
\begin{equation}
\label{eq:ratio}
\rho_{*}(z)= {\cal A}_0 \rho_{\rm BH}(z)
(1+z)^{-\alpha}[1+\lambda_0 (1+z)^{-\beta}].
\end{equation}
Finally, the total star formation rate $\Psi_*$ as a
function of redshift is given by:  
\begin{equation}
 d\rho_{*}(z)/dt = \Psi_{*}(z) - \int_{z_i}^{z}
 \Psi_*(z')\frac{d\chi[\Delta t(z'-z)]}{dt}\frac{dt}{dz'}dz',
\label{eq:sfr}
\end{equation} 
 where $\chi[\Delta t(z'-z)]$ is the fractional mass loss that a
 simple stellar population experiences after a time $\Delta t$
 (corresponding to the redshift interval $(z'-z)$)\footnote{An
 analogous term for \rhobh, due to the ejection of SMBHs from galaxy
 halos after a merger event, is much more difficult to estimate (see
 e.g. Volonteri, Haardt \& Madau 2003) and is neglected here.}. For
 simplicity, we fix $z_i=3.5$, implying that the stellar mass density
 in place then was formed instantaneously at that redshift.  Using the
 year 2000 version of the stellar population synthesis models of
 Bruzual \& Charlot (1993), we have found a very good approximation
 for the function $\chi$ (for $\Delta t \ga 10^{6.5}$ yr) in the form
 $d\chi/dt=(\Delta t)^{-1.17}$. With such a prescription, the
 fractional mass loss after 13 Gyr
 amounts to about 30\%.

\begin{figure}
\psfig{figure=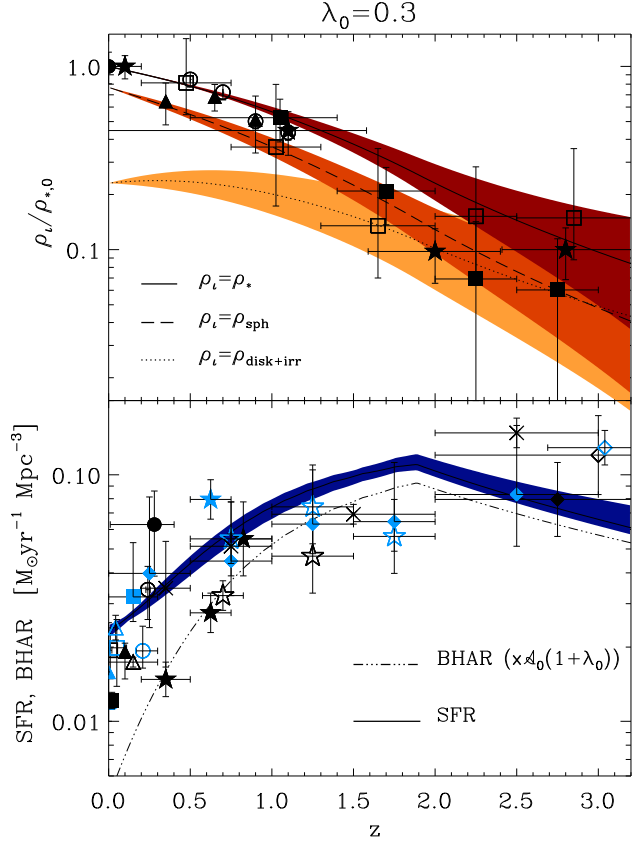,width=0.49\textwidth}
\caption{The top panel shows the evolution in the best fit stellar 
mass density (solid line) as a function of redshift, where the density
is given as a ratio to the local value, $\rho_{*,0} = 5.6 \times
10^8 M_{\odot}$ Mpc$^{-3}$ (Cole et al. 2001).  A value of $\lambda_0=0.3$
and $\rho_{\rm BH,0}=4.2 \times 10^5 M_{\odot}$ Mpc$^{-3}$ (Marconi et
al. 2004) are adopted here. The darker shaded area represents the
1-sigma confidence interval for \rhostar.
 Also shown are the relative decomposition of
the total stellar mass density into \rhosph (dashed line, grey
shaded area) and \rhoelse (dotted line, light grey shaded area), with
their corresponding 1-sigma confidence interval. Please
note that the absolute normalization - and to a lesser degree the
shape - of these two curves depends on the specific value of
$\lambda_0$ adopted (see text for details).  
The data points correspond to measurements from
the 2dFGRS$+$2MASS (Cole et al. 2001 -- filled circle), the Canada
France Redshift Survey (Brinchmann \& Ellis 2000 -- filled triangles),
the MUNICS survey (Drory et al. 2004 -- open circles), the Hubble Deep
Field North (HDF-N; Dickinson et al. 2003 -- filled squares), and the
HDF-S (Fontana et al. 2003 -- open squares; Rudnick et al. 2003 --
filled stars). The lower panel shows the evolution in the best fit SFR
(solid line and dark red shaded area) and the corresponding black hole
accretion rate density (rescaled by a factor ${\cal A}_0
(1+\lambda_0)$).  The data points correspond to the measurements from
Haarsma et al. (2000) -- filled black circles; Condon et al. (2002) --
filled grey circles; Pascual et al. (2001) -- open black circles;
Tresse and Maddox (1998) -- open grey circles; Gallego et al. (1995),
updated following Glazebrook (1999) -- filled black squares;
Sullivan et al. (2000) -- filled grey squares; Serjeant et al. (2002) --
open black squares; Glazebrook et al. (2003) -- open grey
squares; Brinchmann et al. (2004) -- filled black triangles; Gallego et
al. (1995) -- filled grey triangles; Treyer et al. (1998) -- open black
triangles; Gronwall (1998) -- open grey triangles; Lilly et al. (1996)
-- filled black stars; Flores et al. (1999) -- filled grey
stars; Cowie et al. (1996) -- open black stars; Connolly et
al. (1997) -- open grey stars; Madau, Pozzetti and Dickinson
(1998) -- filled black diamonds; Pascarelle, Lanzetta \&
Fernandez-Soto (1998) -- filled grey diamonds; Hughues et al. (1998) --
open black diamonds; Steidel et al. (1999) -- open grey diamonds;
Sawicki et al. (1997) -- black crosses.  }
\label{fig:fits}
\end{figure}

We perform a simultaneous fit to the observed stellar mass density and
SFR points, assuming the functional forms of (\ref{eq:ratio}) and
(\ref{eq:sfr}) above and calculating the black hole mass density
evolution according to eqs.~(\ref{eq:rhobh_z}) and (\ref{eq:bhar_z}).
We vary $\alpha$, $\beta$, and $\epsilon$ over the range $[0,1.5]$,
$[-2,1]$, and $[0.02,0.3]$ respectively.  In Figure~\ref{fig:fits} we
show the observational data for both $\rho_*(z)$ and $\Psi_{*}(z)$,
together with the best fit curves, in this case calculated assuming
$\lambda_0=0.3$ and $\rho_{\rm BH,0} = 4.2 \times 10^5 M_{\odot}$
Mpc$^{-3}$ (Marconi et al. 2004). The shaded areas represent the
1-sigma range of models allowed from the joint fitting of both
datasets.  It is important to note that, for different possible
choices of the local values of $\rho_{\rm BH,0}$ and $\lambda_0$ (see
below), our best fits for \rhostar and $\Psi_*$ are almost unchanged
and always statistically acceptable, with $\chi^2$ values typically
ranging from 43 to 45, for 55 degrees of freedom. Nonetheless, it is
apparent from Fig.~\ref{fig:fits} that the best fit solution for
\rhostar systematically overpredicts the observed points in the
redshift range $1.5\la z\la 2$, where, however, cosmic variance plays
a large role and the \rhostar measurements are quite uncertain.

 Also shown in Fig.~\ref{fig:fits} are the relative contributions to
 the total stellar mass density of \rhosph and \rhoelse. It should
 be stressed that such a decomposition is strongly dependent on
 $\lambda_0$. The {\it relative} growth in spheroids and
 disks~$+$~irregulars relative to their respective local values is in
 fact well represented by these curves, as it mainly depends on the
 value of $\beta$, but their absolute growth, and hence the total mass
 density in disks~$+$~irregulars relative to that in spheroids at any
 $z$, is highly dependent on the adopted value of $\lambda_0$. For
 example, if $\lambda_0=1.0$, the dashed and dotted lines in
 Fig.~\ref{fig:fits} will be moved vertically to attain the same value
 at $z=0$.

\begin{figure}
\psfig{figure=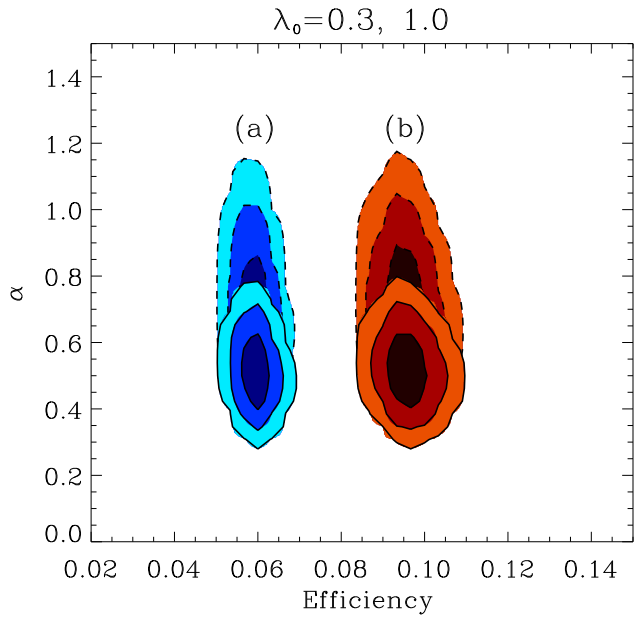,angle=0,width=0.49\textwidth}
\caption{The 1, 2 and 3 sigma confidence levels for
  the parameters $\alpha$ and $\epsilon$ obtained by simultaneously
  fitting the redshift evolution of the total stellar mass density and
  of the total star formation rate density with eqns.
  (\ref{eq:ratio}) and (\ref{eq:sfr}), respectively, marginalizing
  over the other parameter, $\beta$, assuming a flat prior on its
  probability distribution between $-2<\beta<1$. Four sets of contours
  are plotted. Solid lines correspond to a local ratio of the
  disk~$+$~irregular to spheroid stellar mass density of
  $\lambda_0=0.3$; dashed to $\lambda_0=1.0$. Left (a) contours are for
  $\rho_{\rm BH,0}=4.2 \times 10^5 M_{\odot} {\rm Mpc}^{-3}$ (Marconi
  et al. 2004), while the right (b) ones correspond to $\rho_{\rm BH,0}=2.5
  \times 10^5 M_{\odot} {\rm Mpc}^{-3}$ (Yu and Tremaine 2002).}
\label{fig:ef_al}
\end{figure}

 In Figure~\ref{fig:ef_al} we show the results as probability contours
 in the 2-D parameter space $(\epsilon,\alpha)$, obtained by
 marginalizing the total probability distribution over $\beta$,
assuming a flat prior on its distribution, to be conservative. 
The results are shown for two possible values of the
 parameter $\lambda_0$ (0.3 and 1). As it is well known, a constraint
 on the average radiative efficiency of accreting SMBH can be obtained
 by comparing the total mass of the relic population with the
 integrated light from all AGN at all redshifts (Soltan 1982; Yu \&
 Tremaine 2002; Marconi et al. 2004). Given a reliable inventory of
 all the light emitted during the accretion induced growth phase (here
 assumed to be given by the HXLF of AGN), the final result will depend
 crucially on the value of the local SMBH mass density: the lower
 $\rho_{\rm BH,0}$ is, the higher the radiative efficiency has to be
 in order not to over-produce the locally measured density of relic
 black holes.  This is clearly illustrated by Figure~\ref{fig:ef_al},
 that shows the calculated confidence limits assuming two different
 priors for $\rho_{\rm BH,0}$. In one case, shown by the rightmost (b) 
 contours,
 we have adopted a lower value as derived, for example, by Yu \&
 Tremaine (2002) of $\rho_{\rm BH,0}=2.5\times 10^5 M_{\odot} {\rm
 Mpc}^{-3}$. The radiative efficiency is constrained, at the 3-sigma
 level, between $0.08 \la \epsilon \la 0.11$. On the other hand, if
 the local SMBH mass density is as high as $\rho_{\rm BH,0}=4.2 \times
 10^5 M_{\odot} {\rm Mpc}^{-3}$, as suggested by Marconi et
 al. (2004), the radiative efficiency is allowed to take a lower value
 and is constrained, at the 3-sigma level between $0.05 \la \epsilon
 \la 0.07$, as shown by the leftmost (a) contours. In fact, the range of
 allowed radiative efficiencies would have been larger than those
 shown in Fig.~\ref{fig:ef_al} had we also considered the
 uncertainties on the measures of the local black hole mass
 density\footnote{ We note here that the main difference between the
 two values of $\rho_{\rm BH,0}$ comes from a higher normalization of
 the M-$\sigma$ relation by a factor ~1.6 adopted by Marconi et
 al. (2004).  It is beyond the scope of this letter to discuss
 thoroughly the reason of this discrepancy. The reader is referred to
 Yu and Tremaine (2002) and to Marconi et al. (2004) for such a
 discussion.}. It is clear that there is almost complete degeneracy
 between $\epsilon$ and $\rho_{\rm BH,0}$, once the luminosity
 function of AGN is fixed. Efficiencies lower than 0.04 are excluded
 at the more than 4-sigma level, whatever the prior on the local SMBH
 mass density, provided that it is not higher than $6.5\times 10^5
 M_{\odot} {\rm Mpc}^{-3}$. This is easily understood, as black holes
 accreting with $\epsilon<0.04$ build up, in the redshift interval
 between $z=0$ and $z=3$, a mass larger than this value.

\begin{figure}
\psfig{figure=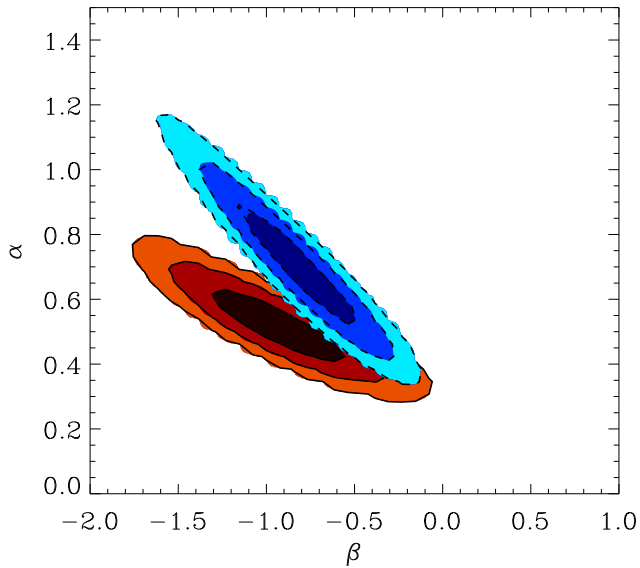,angle=0,width=0.49\textwidth}
\caption{The contours show the 1, 2 and 3 sigma confidence levels for
  the parameters $\alpha$ and $\beta$ obtained by simultaneously
  fitting the redshift evolution of the stellar mass density and of
  the star formation rate density with eqns.~(\ref{eq:ratio}) and
  (\ref{eq:sfr}), respectively, and then marginalizing over the
  radiative efficiency of black holes, $\epsilon$, assuming a flat
  prior on its probability distribution between $0.02<\epsilon<0.4$. 
Solid contours 
  correspond to the case in which we fix the local ratio of
  disk~$+$~irregular to spheroid stellar mass density to
  $\lambda_0=0.3$, dashed contours correspond the the
  case of $\lambda_0=1$. The two parameters are degenerate along the
  line $\alpha+\lambda_0 \beta/(1+\lambda_0) \approx 0.3$ (see text for
  details).}
\label{fig:be_al}
\end{figure}

 Figure~\ref{fig:be_al} shows the probability contours in the 2-D
 parameter space $(\beta,\alpha)$, obtained by marginalizing the total
 probability distribution over $\epsilon$, assuming a flat prior on
 its distribution, once again calculated for two different values of
 $\lambda_0$. We note here that the constraints on these two
 parameters are almost independent of the adopted value for $\rho_{\rm
 BH,0}$. The two sets of contours in figure~\ref{fig:be_al} can be
 understood as follows. At low redshifts, $\Psi_{\rm BH}\ll 1$, and
 $\rho_{\rm BH} \sim const$. Therefore, from 
 eq.~(\ref{eq:ratio}), we see that the observed decline of the total stellar
 mass density, and the corresponding increase of the star formation
 rate density, will be driven by the term $(1+z)^{-\alpha}[1+\lambda_0
 (1+z)^{-\beta}]$. For low values of $z$, this term can be
 approximated by $(1+z)^{-\gamma}$, with $\gamma \simeq \alpha+\beta
 \lambda_0/(1+\lambda_0)$, and fitting the low redshift points (see
 below) with
 this simple expression we obtain the relation $\alpha+\beta
 \lambda_0/(1+\lambda_0)\approx 0.3$, which indeed determines the
 approximate direction along which the best fit parameters $\alpha$
 and $\beta$ are allowed to move.

 As it is expected, the lower the value of $\lambda_0$, the less
 constrained will the parameter $\beta$ be. A constant $\lambda(z)$
 ($\beta=0$) is allowed at the 3-sigma level only if $\lambda_0 <
 0.3$, but the best fit value always correspond to a negative $\beta$,
 i.e. to an increase of the disk/irregulars stellar mass fraction with
 respect to the spheroid one with increasing redshift.
 
  To further test the sensitivity of our constraints on $\alpha$,
   $\beta$ and $\epsilon$ we have also performed a fit restricted to
   the redshift interval $0 \le z \le 1$, where the observational data
   points are more numerous and have smaller error bars.  In doing
   this, we have found that while the constraints on $\alpha$ and
   $\beta$ remain virtually unchanged, those on the radiative
   efficiency are much looser. In particular, for high values of the
   local SMBH mass density, we find that  $0.03 \la \epsilon \la 0.12$ at
   the 3-sigma level. This can be understood because the constraint on
   $\epsilon$ comes primarily from the epoch in which the black hole
   accretion rate density is highest, which occurs at $z>1$, where the
   QSO luminosity function peaks.  The data at $z<1$ don't serve to
   strongly constrain $\epsilon$.
   
   In summary, irrespective of all the uncertainties in the determination of
   $\lambda_0$ and in the high redshift stellar mass and star
     formation rate points, the data do indeed put some constraints
   on the redshift evolution of $\rho_{\rm sph}/\rho_{\rm BH}$.  In
   particular, a constant ratio, $\alpha=0$ is excluded at a more than
   4-sigma confidence level, implying that the black hole to spheroid
   mass density ratio must have been larger in the past.

\section{Discussion}
It is well known that both AGN activity and SFR decline at low
redshift. Moreover, the correlation observed locally between
supermassive black hole masses and global properties of their host
spheroids are suggestive of a fundamental link between the formation and
evolution of black holes and galaxies.

In this letter we have attempted to constrain phenomenologically the
joint evolution of supermassive black holes, their host spheroids and
the total stellar mass density in the universe.

 We have assumed that the black hole mass density evolution is due to
 accretion only and that it can therefore be reliably reconstructed
 from the local black hole mass density and from the redshift
 evolution of the hard X-ray luminosity function of AGN/QSO, with the
 (average) radiative efficiency of accretion, $\epsilon$ as the only
 free parameter. We have then examined the relation between $\rho_{\rm
 BH}$ and $\rho_{\rm sph}$, assuming that the ratio of these two
 densities simply evolves as $(1+z)^{-\alpha}$. Finally, we have
 considered the simple case in which the ratio of the stellar mass
 density in disk components~$+$~irregular galaxies to that of the
 spheroid component of galaxies evolves as $\lambda(z)\propto
 (1+z)^{-\beta}$.  By fitting simultaneously all the available
 observational data on the total stellar mass density and star
 formation rate density as a function of redshift, we have obtained a
 set of constrains on the three model parameters
 $(\epsilon,\alpha,\beta)$, which themselves depend on the local
 values of $\rho_{\rm BH}$ and $\lambda$.

 We have shown that the constraint we obtain on the radiative
 efficiency of accretion crucially depends on the adopted value for
 the local black hole mass density, but not on $\alpha$ or $\beta$.

 More importantly, we have shown that BH accretion does not exactly
 track either the spheroid nor the total star assembly (i.e. both
 $\alpha$ and $\gamma=\alpha+\lambda_0 \beta/(1+\lambda_0)$ are larger
 than zero). This is the most important result of our study:
 irrespective of the exact mass budget in spheroids and
 disks~$+$~irregulars, the ratio of the total or spheroid stellar to
 black hole mass density was lower at higher redshift.  The
 simultaneous fit we obtain for both $\rho_{*}$ and $\Psi_{*}$also
 indicates that the black hole growth rate is suppressed with respect
 to the total SFR at $z < 2$, as it can be seen in
 Figure~\ref{fig:fits} (as also shown in the cosmological simulations
 of Di Matteo et al. 2003).

 Our findings have a direct {\it observable} consequences for the
 expected redshift evolution of the $M_{\rm BH}-M_{\rm sph}$
 relationship: $\alpha>0$ implies a larger black hole mass for a given
 host spheroid mass at higher redshift.

 We can also compare our results with recent semi-analytical work that
 tries to incorporate SMBH growth and feedback into a galaxy evolution
 scenario. The simplest way to relate black hole mass and global
 galactic properties is via an energy argument, as discussed, for
 example, in Wyithe and Loeb (2003): a SMBH will stop growing as soon
 as the energy provided by accretion onto it at a fraction $\eta$ of
 the Eddington rate for a time $t_{\rm Q}$ will equal the binding
 energy of the gas in the DM halo: $M_{\rm BH}\eta L_{{\rm Edd},\odot}
 t_{\rm Q} F_{\rm Q}=\frac{1}{2}{\Omega_{\rm b}}{\Omega_{\rm m}}
 M_{\rm halo} v_{\rm c}^2.$ Here, the black hole mass is in solar
 units, $L_{{\rm Edd},\odot}$ is the Eddington rate of a one solar
 mass black hole, $F_{\rm Q}$ is a coupling coefficient that can be
 determined from the normalization of the $M_{\rm BH}-\sigma$
 relation, $\Omega_{\rm m}$ and $\Omega_{\rm b}$ are the matter and
 baryon densities, respectively, and $M_{\rm halo}$ $v_{\rm c}$ are
 the mass and circular velocity of a DM halo.  Most of the feedback
 models in the recent literature (Kauffmann and Haenelt 2000;
 Cavaliere \& Vittorini 2002; Menci et al. 2003; Wyithe and Loeb 2003;
 Granato et al. 2004) assume (sometimes implicitly) a linear
 proportionality between the quasar lifetime and the dynamical time of
 the dark matter halo in which it is embedded: $t_{\rm Q}=t_{\rm
 dyn}\propto \xi(z)^{-1/2} (1+z)^{-3/2}$, where $\xi(z)$ is a weak
 function of $z$ and depends on the cosmological parameters only (see
 e.g.; Barkana and Loeb 2001 for more details). Indeed, Wyithe and
 Loeb (2003) have shown that, if the QSO lifetime is equal to the
 dynamical time of a DM halo, then the $M_{\rm BH}-\sigma$ relation
 does not change with redshift, while the ratio of the black hole to
 (spheroid) stellar mass density should evolve as $\rho_{\rm
 sph}/\rho_{\rm BH}\propto [\xi(z)]^{-1/2}(1+z)^{-3/2}\sim
 (1+z)^{-1.15}$, where the final approximation is valid for $0<z<2$.
 Comparing this with the constraints on $\alpha$ we have obtained in
 section~\ref{sec:results}, we see that, although marginally consistent with
 these simple semi-analytical schemes, and only for the high
 $\lambda_0$ case, the data clearly suggest a slower evolution for the
 spheroid to black hole mass ratio than that predicted by just
 assuming a constant $M_{\rm BH}-\sigma$ relation at all
 $z$. Therefore, this would imply that the normalization of the
 $M_{\rm BH}-\sigma$ relation {\it should evolve} with redshift.  A
 caveat is however in place here, as of course 
 $\eta$ and $F_{\rm Q}$ in prescriptions such as those of
 Wyithe and Loeb (2003) may not be constant, but vary as
 a function of black hole mass, or mass accretion and as
 function of redshift (e.g.; Merloni 2004).
 Observational tests of the $M_{\rm BH}-\sigma$ and 
 $M_{\rm BH}-M_{\rm sph}$ relations at higher redshift (see e.g. Shields et
 al. 2003) will be crucial to our understanding of the interplay
 between black holes and their host galaxies.
 
 Finally, our results suggest that the fraction of \rhostar locked up
 into the non-spheroidal components of galaxies and in irregular
 galaxies should increase with increasing redshift. This result is
 consistent with Brinchmann \& Ellis (2001) that show that the mass
 density in irregulars increases rapidly out to $z\sim 1$.
 
 If we relax the assumption that black hole mass and spheroid
 mass densities are related at all redshifts, then \rhosph and \rhoelse
 in eq.~(\ref{eq:lambda}) would simple correspond to the stellar mass
 that {\it ends up} in spheroids and disks, respectively, at $z=0$,
 where we know that black hole mass and spheroid mass are related.
 Under this assumption, negative values of $\beta$ would imply that
 the stars in local spheroids were formed at a later time than the
 stars in present day disks. This would directly contradict the
 relative observed ages of disks and spheroids.  On the other hand, no
 such contradiction occurs when using our adopted assumption,
 i.e. that there is a $M_{\rm BH}-M_{\rm sph}$ relation at all
 redshifts and that \rhosph and \rhoelse correspond, at any given
 redshift, to the mass in spheroids and disks~$+$~irregulars
 respectively. This further supports the idea that the dynamical
 events that lead to the assembly of spheroid stellar mass (mergers,
 secular instabilities, gas accretion, etc.) are directly linked to
 the major accretion events onto the SMBH.
 
 The upcoming results from projects such as the GEMS (Galaxy Evolution
 from Morphological Studies; see Rix et al. 2004) survey, that will
 provide morphologies and structural parameters for nearly 10,000
 galaxies at $0.2\la z \la 1.1$ and their respective contributions to
 the mass density, will surely improve our understanding of this
 issue, and will possibly help us to better understand how the
 simultaneous evolution of black holes, spheroids, and disks in
 galaxies proceeded.

\section*{Acknowledgments}
We thank J. Brinchmann, A. Fontana, and V. Springel for providing us
with some of the observational data on star formation and stellar mass
density.  G.R. acknowledges the support by the Deutsche
Forschunggemeinschaft (DFG), SFB 375 (Astroteilchenphysik).

\bsp

\label{lastpage}

\end{document}